\documentclass[aps,prl,reprint,longbibliography,notitlepage,superscriptaddress,preprintnumbers]{revtex4-1}

\usepackage{graphics}      % standard graphics specifications
\usepackage{graphicx}      % alternative graphics specifications
\usepackage{url}           % for on-line citations
\usepackage{bm}            % special 'bm-math' package
\usepackage{amsmath}
\usepackage{amssymb}
\usepackage{physics}
\usepackage[caption=false]{subfig}

\usepackage{capt-of}%%To get the caption

\usepackage[colorlinks=true, pdfstartview=FitV, linkcolor=blue, citecolor=blue, urlcolor=blue]{hyperref}

\newcommand{\dhd}{{\textstyle d}
	\lower.03ex\hbox{\kern-0.38em$^{\scriptstyle-}$}\kern-0.05em{}}
\newcommand{\dbar}{{\textstyle \delta}
	\lower.03ex\hbox{\kern-0.38em$^{\scriptstyle-}$}\kern-0.05em{}}
\newcommand{\half}{{1\over 2}}

\newcommand{\baru}{{\bar u}}

\newcommand{\calm}{{\cal M}}

\newcommand{\calu}{{\cal U}} 
\newcommand{\calv}{{\cal V}}

\newcommand{\barpsi}{{\bar \psi}}

\newcommand \sslash [1] {\slash\hspace{-0.2cm}{#1}}

\newcommand{\ssx}{\sslash{x}}
\newcommand{\ssy}{\sslash{y}}
\newcommand{\ssz}{\sslash{z}}

\newcommand \Slash [1] {\slash\hspace{-0.23cm}{#1}}

\newcommand{\Sx}{\Slash{X}}
\newcommand{\Sy}{\sslash{Y}}

\begin{document}
	
	\title{Small-x behavior of quark pseudo- and quasi-PDFs}
	
	\author{Giovanni Antonio Chirilli}
	\email{giovanni.chirilli@ur.de}
	\affiliation{Institut f\"ur Theoretische Physik, Universit\"at Regensburg,\\
		Universit\"atsstrasse 31, D-93040 Regensburg, Germany}
	\begin{abstract}
		
The pseudo- and the quasi-PDFs defined through space-like bilocal operators allow direct access 
to the parton distribution functions from first principles in lattice gauge theory.
This formalism, however, leaves the small Bjorken $x$ regime inaccessible. In view of the future Electron-Ion Collider, the study of the PDFs at small-$x$ is timely. I will calculate the quark pseudo- and quasi-PDF in the high-energy limit and show that they have rather different behavior.
In particular, I will show that the quark pseudo-PDF has the expected small-$x$ behavior while the quasi-PDF becomes ill-defined at this regime.

	\end{abstract}
\maketitle 
	
{\it Introduction} --- 
In the last decade, the study of euclidean separated, gauge invariant, bi-local operators through 
lattice gauge formalism has dragged a lot of attention because it gives direct access to the 
parton distribution functions (PDFs) from first principles.
The idea that ignited such a surge of activity is based on
the observation~\cite{Ji:2013dva} that space-like separated operators can be studied with 
lattice QCD formalism and that, in the infinite momentum frame, they reduce to the usual light-cone operators through which the 
parton distribution functions (PDF) are defined. Deviation from the infinite momentum frame come in as inverse powers of the 
large parameter of the boost, thus, it was thought that such corrections can be systematically suppressed. 

The PDFs introduced in Ref.~\cite{Ji:2013dva} are known as quasi-PDFs.
Later, in Ref.~\cite{Radyushkin:2017cyf}, alternative PDFs were introduced known as pseudo-PDFs.
The Bjorken $x$ ($x_B$) dependence of these two distributions has been intensively studied 
in the recent years~\cite{Chen:2016utp, Orginos:2017kos, Ji:2017oey, HadStruc:2021wmh}, 
but, unfortunately, lattice formalism does not allow access to their behavior at small $x_B$ values. 
In view of the future QCD colliders, like the Electron-Ion Collider~\cite{Accardi:2012qut} (EIC) in the USA,
the EIC in China~\cite{Anderle:2021wcy}, and the Large Hadron electron Collider~\cite{LHeC:2020van} (LHeC) in Europe,
where the small-$x_B$ dynamics will be probed, knowledge of the pseudo- and quasi-PDF at this regime is timely.

In this work, we will show that, although defined through the same space-like separated bilocal operators,
the quark quasi-PDFs and pseudo-PDFs have rather different behavior at small$x_B$. A similar conclusion was recently reached also
for the gluon case~\cite{Chirilli:2021euj}.

The small-$x_B$ behavior of the deep inelastic scattering (DIS) structure function can be calculated through the 
high-energy operator product expansion (OPE)~\cite{Balitsky:1995ub} 
where the T-product of two electromagnetic current, ${\rm T}\{J^\mu(x)J^\nu(y)\}$,
is expanded in terms of the coefficient functions, known also as impact factors, convoluted with the matrix elements of infinite Wilson lines.
The evolution equation of the matrix elements of infinite Wilson lines
with respect to the rapidity factorization parameter, is the BK equation~\cite{Balitsky:1995ub,Kovchegov:1999yj}, 
which, with its linear term, takes care of the small-$x_B$ leading-log resummation,
and with its non-linear one, takes care of the unitarity property of the theory. 
Because of its nonlinear nature, the BK equation is regarded as a generalization of the BFKL equation.

The high-energy OPE is the sister procedure of the OPE in non-local (finite gauge-link) operators
adopted in the Bjorken limit~\cite{Balitsky:1987bk} where the factorization scale is the transverse momentum of the fields, and the
evolution equation of the finite gauge-link with respect to the factorization parameter is the DGLAP evolution equation.

In the DGLAP regime, the structure functions of DIS are governed by the anomalous dimensions of twist-2
operators. On the other hand, in the BFKL regime, the structure functions receive contributions from the infinite series of twist expansion.
Relaxing the BFKL limit to enter the overlapping region, from the small-$x_B$ structure function one may obtain the anomalous dimension
of twist-2 (and presumably higher twist) operators in the small-$x_B$ limit. Indeed, in ref.~\cite{Fadin:1998py},
using the next-to-leading (NLO) order BFKL equation (the NLO pomeron intercept),
the small-$x_B$ limit of twist-2 anomalous dimensions at the next-to-next leading order was predicted. 
With the analytic continuation of the twist-2 local operator to non-integer spin, the local operators become a non-local
light-ray operators through which the prediction of the small-$x_B$ limit of the 
twist-2 anomalous dimension from the BFKL resummation becomes more transparent~\cite{Balitsky:2014sqe, Balitsky:2013npa}.
As we will show in this work, this formalism will allow us to obtain the leading and next-to-leading twist corrections from the full BFKL
resummed results for the pseudo- and quasi-PDF.

{\it Pseudo Ioffe-time distribution at high-energy} ---
The quark distribution is defined through the light-cone matrix element~\cite{Balitsky:1987bk}:
\begin{eqnarray}
&&\hspace{-5mm}
\langle P|\bar\psi(z)\ssz[z,0]\psi(0)|P\rangle
=2P\!\cdot\! z\!\!\int_{-1}^1\!\! dx_B\,e^{iP\cdot z\, x_B}Q(x_B)
\label{def-qpdf2}
\end{eqnarray}
with $z^2=0$, $P$ the hadronic momentum, and with the gauge link defined as
\begin{eqnarray}
\hspace{-0.3cm}
[x,y] = {\rm P}{\rm exp}\left\{ig\!\!\int_0^1\!\! du\,(x-y)^\mu A_\mu \left(x+(1-u)y\right)\right\}\,.
\end{eqnarray} 
The distribution $Q(x_B)$ is defined through the quark distribution, $D_q$, and anti-quark distribution, $D_{\bar{q}}$ as
\begin{eqnarray}
Q(x_B)~=~\theta(x_B)D_q(x_B)-\theta(-x_B)D_{\bar{q}}(-x_B)\,.
\end{eqnarray}

To study the small-$x_B$ behavior of the pseudo- and quasi-PDF,
we consider the structure functions defined through the quark pseudo-Ioffe-time distribution (pseudo-ITD)~\cite{Radyushkin:2017cyf}
\begin{eqnarray}
\calm(\nu,z^2) = {z_\mu\over 2\nu}\bra{P}\barpsi(z)\gamma^\mu[z, 0]\psi(0)\ket{P}
\label{defoperator}
\end{eqnarray}
where we defined the Ioffe-time parameter $\nu=z\cdot P$.
The pseudo-ITD, (\ref{defoperator}), is a bi-local operator with space-like separation $z^2<0$.

To study the high-energy behavior of the pseudo-ITD, we perform an infinite longitudinal boost in
coordinates space. In this limit, we do not distinguish between the
$0$-th, and the $3$-rd components, and we can rewrite the pseudo-ITD (\ref{defoperator}) as~\cite{Chirilli:2021euj}
\begin{eqnarray}
\hspace{-0.2cm}&&\bra{P}\barpsi(L,x_\perp)\gamma^-[Ln^\mu+x_\perp, 0]\psi(0)\ket{P + \epsilon^- n'}\,2\pi\delta(\epsilon^-)
	\nonumber\\
	&& = \int_0^{+\infty} \!\!dx^+\!\! \int_{-\infty}^0\!\!dy^+\delta(x^+-y^+-L)\bra{P}\barpsi(x^+,x_\perp)\gamma^-
	\nonumber\\
	&&~~\times[x^+n^\mu+x_\perp, y^+n^\mu+0_\perp]\psi(y^+,0_\perp)\ket{P+\epsilon^-n'}\,.
	\label{IoffeAmpOperator}
\end{eqnarray}
In eq. (\ref{IoffeAmpOperator}) we introduced the light-cone vectors $n$ and $n'$
such that $n^2=n'^2=0$, $n\cdot n'=1$, and $x\cdot n'=x^+$, $x\cdot n =x^-$, where $x^\pm={x^0\pm x^3\over \sqrt{2}}$.
In this way, a generic vector is written as $x^\mu=x^+n^\mu + x^- n'^\mu + x^\mu_\perp$, where
the transverse coordinates are $x^\mu_\perp = (0,x^1,x^2,0)$.
Here, the length of the gauge link is indicated by the real parameter $L$.

Our goal is to study the operator (\ref{IoffeAmpOperator}) in the high-energy limit, and
obtain the small-$x_B$ behavior of the pseudo- and quasi-PDF.
The same procedure was adopted in \cite{Chirilli:2021euj} where we calculated
the high-energy behavior of the gluon pseudo-ITD and where it was shown that
the high-energy behavior is obtained by resumming large logarithms of the Ioffe-time parameter $\nu$,
and that such resummation is obtained by the solution of the evolution equation with the Ioffe-time as evolution parameter.

The high-energy OPE is a well-established formalism to study the high-energy behavior of structure functions in coordinate space.
In Ref.~\cite{Balitsky:2010ze,Balitsky:2012bs}, this formalism was used to obtained the next-to-leading (NLO) structure function for
DIS, while in Ref.~\cite{Balitsky:2009xg} it was used to study the high-energy behavior of the 
four-point function in the super symmetric $\cal N$=4 Yang-Mills theory. 

\begin{figure}[htb]
	\begin{center}
		\includegraphics[width=2.3in]{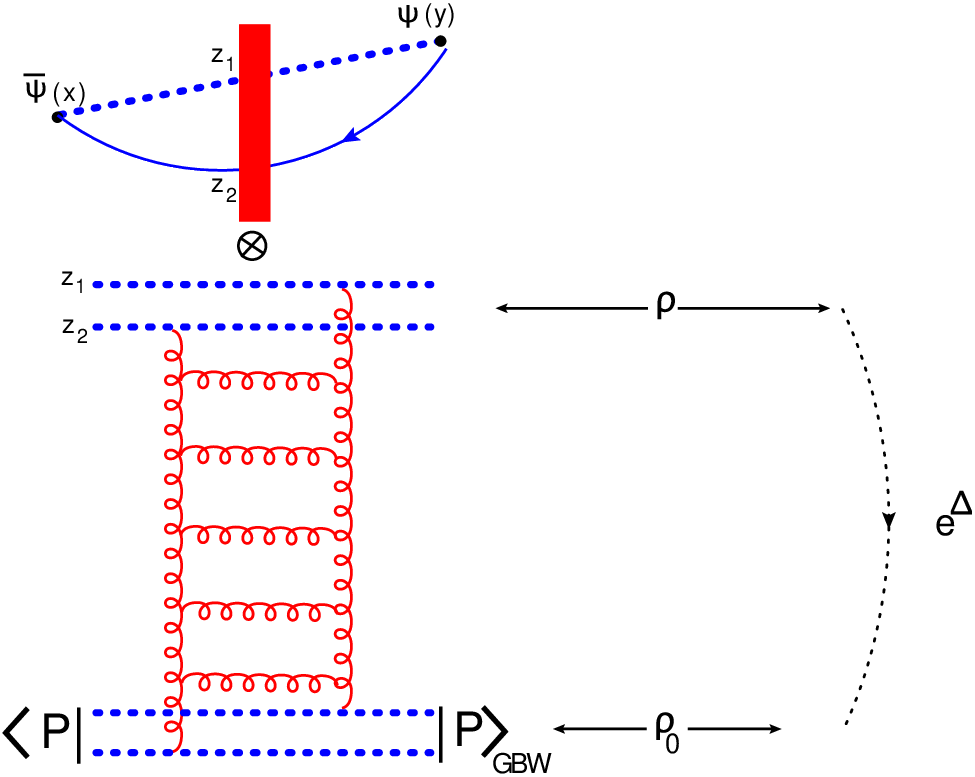}
		\caption{Diagram for the LO impact factor.
			We indicate in blue the quantum fields and in red the classical background ones.}
		\label{Fig:quarkLOif-evolution}
	\end{center}
\end{figure}

We start calculating the impact factor which is the top part of
the diagram in Fig.\ref{Fig:quarkLOif-evolution}. To this end, we consider 
the operator in eq. (\ref{IoffeAmpOperator}) in the gluon external field which, at high-energy shrink to a shock-wave
(red vertical band in the figure) because of Lorentz contraction and time-dilation. 
We need the quark propagator in the shock-wave in coordinate space\cite{Balitsky:2001gj}
\begin{eqnarray}
\hspace{-0.3cm}	\langle\psi(x)\barpsi(y)\rangle \stackrel{y^+<0<x^+}{=} 
	&&\int d^4z\delta(z^+){\ssx-\ssz\over 2\pi^2[(x-z)^2-i\epsilon]^2}
	\nonumber\\
	&&\times\sslash{n}'U_z{\ssy-\ssz\over 2\pi^2[(y-z)^2-i\epsilon]^2}\,.
	\label{LO-DIS-OPE}
\end{eqnarray}	
where $U_x = U(x_\perp) = {\rm P}{\rm exp}\left\{ig\int_{-\infty}^{+\infty} dx^+A^-(x^+,x_\perp)\right\}$
is the infinite Wilson line. 
In the propagator (\ref{LO-DIS-OPE}), the shock-wave is located at the $0^+$ coordinate point, and we are considering
the case in which the quark starts its propagation at $y^+<0$, goes through the background of shock-wave at the
point $z^+=0$, and ends its propagation at $x^+>0$. In eq. (\ref{LO-DIS-OPE}), we signaled this
with the superscript $y^+<0<x^+$.

Functionally integrating over the quantum quark field, the impact factor is given by the product of two quark propagators
(\ref{LO-DIS-OPE})
\begin{eqnarray}
	&&\hspace{-0.1cm}
	\int dx^+ dy^+\delta(x^+-y^+-L)
	\nonumber\\
	&&\hspace{-0.1cm}
	\langle\barpsi(x^+,x_\perp)\gamma^-[x^+n^\mu+x_\perp, y^+n^\mu+y_\perp]\psi(y^+,y_\perp)\rangle_A
	\nonumber\\
	&&\hspace{-0.1cm}
	\stackrel{y^+<0<x^+}{=}  \int_0^{+\infty} {dx^+\over {x^+}^2} \int_{-\infty}^0d{y^+\over {y^+}^2}\, \delta(x^+-y^+-L)
	\nonumber\\
	&&\hspace{-0.1cm}
	\times\!\!\int {d^2z_2\over 2\pi^3}
	{- 4\,i\,(X_2,Y_2)_\perp\over \Big[{Y^2_{2\perp}\over |y^+|} + {X^2_{2\perp}\over x^+}+i\epsilon\Big]^3}
	\langle \Tr\{U_{z_1}U^\dagger_{z_2}\}\rangle_A\,,
	\label{quark-startinpoint}
\end{eqnarray} 
where we used the short-hand notation $X_{1\perp} = x_\perp-z_{1\perp}$ and $Y_{1\perp}=y_\perp-z_{1\perp}$,
and also used $\tr\{\gamma^-\Sy_{2\perp}\gamma^+\Sx_{2\perp}\}= 4(X_2,Y_2)_\perp$
with the two dimensional scalar product defined as $(x,y)_\perp = x^1y^1+x^2y^2$.
The subscript $A$ on the angle bracket means that the operator is being evaluated in the background of the gluon field.

Note that we indicated with $z_2$ the point at which the shock wave cut the quark propagator.
The point $z_1$, instead, is the point at which the shock wave cut the straight dotted blue line which represents the gauge link, and it is a
point that runs from point $x$ to point $y$. 

The evolution of the trace of two infinite Wilson lines with respect to the rapidity is the BK equation. However,
for our purpose, it will be sufficient considering only the linearization of the BK eq. i.e. the BFKL equation which is
\begin{eqnarray}
&&\hspace{-1.5cm}2a{d\over da}\,\calv^a(z_\perp) 	
= {\alpha_s N_c\over \pi^2}\!\int\!d^2z'
\nonumber\\
&&\times
\Big[{\calv^a(z'_\perp)\over (z-z')_\perp^2} - {(z,z')_\perp\calv^a(z_\perp)\over z'^2_\perp(z-z')^2_\perp}\Big]\,,
\label{calvevolution}
\end{eqnarray}
with ${1\over z^2_\perp}\calu(z_\perp) \equiv \calv(z_\perp)$,
and with  $\calu(z_\perp)$ the forward matrix element which depends only on its transverse size and it is obtained from
$\calu(x_\perp,y_\perp) = 1-{1\over N_c}\tr\{U(x_\perp)$ $U^\dagger(y_\perp)\}$.

Equation (\ref{calvevolution}) is an evolution equation with respect to the coordinate space parameter defined as
\begin{eqnarray}
a = -{2x^+y^+\over (x-y)^2a_0}+i\epsilon\,.
\label{coordparam}
\end{eqnarray}
which is reminiscent of the composite Wilson lines operators introduced to restore the 
M\"obius conformal invariance lost at NLO level (for details see Refs. 
\cite{Balitsky:2009xg, Balitsky:2009yp, Balitsky:2010ze, Balitsky:2012bs}).
The peculiarity of the evolution parameter $a$ is that it is in coordinate space and it suits well our purpose.

The solution of evolution equation (\ref{calvevolution}) is
\begin{eqnarray}
\calv^a(z_{12}) = && -i\int_{\half-i\infty}^{\half+i\infty}\!{d\gamma\over 2\pi^2}(z_{12}^2)^{\gamma-1} 
\left(a\over a_0\right)^{\aleph(\gamma)\over 2}
\nonumber\\
&&\times\int d^2\omega(\omega^2_\perp)^{-\gamma}\calv^{a_0}(\omega_\perp)\,,
\label{solution}
\end{eqnarray}
where $\aleph(\gamma)\equiv \bar{\alpha}_s\chi(\gamma)$, with $\bar{\alpha}_s = {\alpha_s N_c\over \pi}$,
$\chi(\gamma)=  2\psi(1) - \psi(\gamma) - \psi(1-\gamma)$.
The parameter $a_0$ is the initial point of evolution  which is $a_0={P^-\over M_N}$ with $M_N$ the mass of the hadronic target
and $P^- = P\!\cdot\! n$. 

We will use notation  ${Y^2_\perp\over |y^+|} + {X^2_\perp\over x^+}
= {1\over \Delta^+u\baru}\left[(z_2-x_u)^2_\perp + \Delta^2_\perp u\baru\right]$
with $\Delta^+ = x^++|y^+| = L$, $u={|y^+|\over \Delta^+}$, $\baru = {x^+\over \Delta^+}$, and 
$x_u = ux_\perp + \baru y_\perp$. Moreover, we observe that, $z_{1\perp}$ is a point which run from $x_\perp$ to $y_\perp$, so
$x_u=z_{1\perp}$. 
With these new notations the solution of the BFKL equation, eq. (\ref{solution}), becomes
\begin{eqnarray}
&&\hspace{-0.5cm}\calv^a(z_{12})
=  -i\int_{\half-i\infty}^{\half+i\infty}{d\gamma\over  2\pi^2}
	\,(z_{12}^2)^{\gamma-1}
	\nonumber\\
	&&\hspace{-0.5cm}\times
	\!\!\int d^2\omega (\omega^2_\perp)^{-\gamma}\calv^{a_0}(\omega_\perp)
	\left(-{2L^2u\baru\over \Delta_\perp^2}{{P^-}^2\over M^2_N}+i\epsilon\right)^{{\aleph(\gamma)\over 2}}
	\label{solution2}
\end{eqnarray}
with $\calv^a(z_{12}) = {1\over z_{12}^2}\calu^a(z_{12})$, and $\Delta = (x-y)$. 
As mentioned above, we will use the GBW model~\cite{Golec-Biernat:1998js}
(alternatively one could use the MV model~\cite{McLerran:1993ni}) for the forward dipole matrix element
\begin{eqnarray}
&&\bra{P}\calu(x)\ket{P+\epsilon^-n'}
\nonumber\\
~~=&&P^-2\pi\delta(\epsilon^-) \sigma_0\left(1-\exp\left({-x^2_\perp Q^2_s\over 4}\right)\right)
\label{GBWmodel}
\end{eqnarray}
So, using model (\ref{GBWmodel}) in eq. (\ref{solution2}) and integrating over $\omega_\perp$, from eq. (\ref{quark-startinpoint})
 we arrive at
\begin{eqnarray}
&&\hspace{-1.2cm}\calm(\nu,z^2)
 = {N_c\sigma_0\over 2\pi|z|^2}\int_{\half-i\infty}^{\half+i\infty} d\gamma
	\left({2\nu^2\over z^2M^2_N}+i\epsilon\right)^{{\aleph(\gamma)\over 2}}
	\nonumber\\
	&&\vspace{0.3cm}\times
	{\gamma^3\over \sin^2(\pi\gamma)}{\Gamma(\gamma)\over \Gamma(2+2\gamma)}
	\left({Q^2_s|z|^2\over 4}\right)^\gamma\,,
	\label{quark-bilocalresult-z}
\end{eqnarray}
where we made use of the space like vector $z^\mu$, $z^2<0$, and, in the high-energy limit, we identify
the Ioffe-time parameter $\nu=z\!\cdot\! P= LP^-$ with $P^-= {z^\mu \over|z|}P_\mu$, and 
$\gamma^-= {z^\mu \over|z|}\gamma_\mu$.

\begin{figure}[htb]
	\begin{center}
		\includegraphics[width=1.6in]{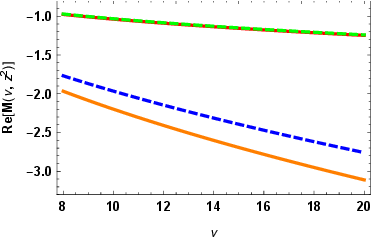}
		\includegraphics[width=1.6in]{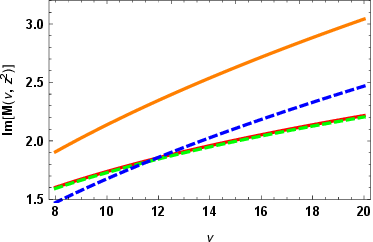}
		\caption{In the left and right panels we plot the real and imaginary parts, respectively, of the quark pseudo-ITD. We compare 
			full BFKL resummation given by 
			the numerical evaluation of eq. (\ref{quark-bilocalresult-z}) 
			(orange curve) with its saddle point approximation, eq. (\ref{q-saddpoint}). For comparison, we also plot
			the LT, (green dashed curve), and the NLT (red solid curve) given in eq. (\ref{LT-NLTioffeamp}).}
		\label{Fig:IofAm-twistSadBFKLReIm}
	\end{center}
\end{figure}

The evaluation of the integral over $\gamma$ in result (\ref{quark-bilocalresult-z}) can be done numerically or
in the saddle point approximation.
In eq. (\ref{quark-bilocalresult-z}),  
the factor ${\gamma^3\over \sin^2(\pi\gamma)}{\Gamma(\gamma)\over \Gamma(2+2\gamma)}$ 
is a slowly varying function. Thus, in the saddle point approximation, eq. (\ref{quark-bilocalresult-z}) is
\begin{eqnarray}
&&\hspace{-0.7cm}
\calm(\nu,z^2)
\simeq {i\,N_c\over 64}{Q_s\sigma_0\over|z|}
{e^{-{\ln^2 {Q_s|z|\over 2}\over 7\zeta(3)\bar{\alpha}_s\ln\left({2\nu^2\over z^2 M^2_N}+i\epsilon\right)}}
	\over\sqrt{7\zeta(3)\bar{\alpha}_s\ln\left({2\nu^2\over z^2 M^2_N}+i\epsilon\right)}}
\nonumber\\
&&\hspace{0.5cm}\times\!\left({2\nu^2\over z^2 M^2_N}+i\epsilon\right)^{\bar{\alpha}_s 2\ln 2}\,.
\label{q-saddpoint}
\end{eqnarray}
In Fig. \ref{Fig:IofAm-twistSadBFKLReIm} we plotted the real and the imaginary parts of the quark Ioffe-time distribution in the saddle point approximation,
eq. (\ref{q-saddpoint}), (blue dashed curve), and the numerical evaluation of eq. (\ref{quark-bilocalresult-z}) (orange curve). 
The plots were obtained using $Q_s=(x_0/x_B)^{0.277\over 2}$ with $x_0=0.41\times 10^{-4}$~\cite{Golec-Biernat:1998js}.
However, for the Ioffe-time amplitude, we will use $Q_s=(x_0\nu_0)^{0.277\over 2}$ where $\nu_0$
is the starting point of the evolution, which, as can be observed from Fig. \ref{Fig:IofAm-twistSadBFKLReIm}, is $\nu_0=8$,
so $Q_s=0.33 \, {\rm GeV}$. The value of the strong coupling we used is $\bar{\alpha}_s = {N_c\alpha_s\over \pi} = 0.2$.

{\it Pseudo Ioffe-time distribution at leading and next-leading twist} ---
In this section, we will calculate the leading twist (LT) and next-to-leading twist (NLT) contribution to the Ioffe-time distribution.
This procedure was already used in Ref.~\cite{Chirilli:2021euj}, and, as mentioned in the Introduction,
relies on the analytic continuation of local operators to non-integer complex spin, thus allowing to the analysis of the scattering
amplitude in the complex plane. Note that this technique is similar to the complexification of the angular momentum 
in the partial wave amplitude in Regge theory.

We start from eq. (\ref{quark-bilocalresult-z}), take its  Mellin transform and obtain
\begin{eqnarray}
	&&\int_{\Delta^2_\perp M_N}^{+\infty}dL\, L^{-j}\int_0^{+\infty} dx^+ \int_{-\infty}^0dy^+\delta(x^+-y^+-L)
	\nonumber\\
	&&\times\langle\barpsi(x^+,x_\perp)\gamma^-[x^+n^\mu+x_\perp, y^+n^\mu+y_\perp]\psi(y^+,y_\perp)\rangle
	\nonumber\\
	&&={N_c\over \pi^3\Delta^2_\perp}\int_{\half-i\infty}^{\half+i\infty} \!\! d\gamma\,
	{\theta\big(\Re[\omega - \aleph(\gamma)]\big)}
	{(\Delta^2_\perp M_N)^{-\omega + \aleph(\gamma)}\over \omega - \aleph(\gamma)}
	\nonumber\\
	&&\times\left({Q^2_s\Delta^2_\perp\over 4}\right)^{\!\!\gamma}
	{\Gamma^2(1-\gamma)\Gamma^3(1+\gamma)\over\Gamma(2+2\gamma)}
	\,\left(\!-{2\over \Delta_\perp^2}{{P^-}^2\over M^2_N}+i\epsilon\right)^{\!\!{N(\gamma)\over 2}}\,.
	\label{quarkMellin1}
\end{eqnarray}
One should notice that the analytic continuation of local twist-2 operators to complex values of the spin, thus generating
non-local operators (see Refs.~\cite{Balitsky:2018irv, Chirilli:2021euj} for details), is indeed equivalent to taking 
the Mellin transform of the non-local operator in eq. (\ref{quarkMellin1}).
The integration over $\gamma$ can be calculated by closing the contour to the right and taking the residue.
Taking the residue at the value of $\tilde{\gamma}$ such that $\omega-\aleph(\tilde{\gamma})=0$
provides the exact BFKL result as resummation of an infinite series of, using Regge theory terminology, \textit{moving} poles.
However, we will take such moving poles one by one and show that they organize themselves as an expansion, the twist expansion,
where each term is suppressed as powers of $x^2_\perp$ for $x^2_\perp\to 0$.  

Now, let us observe that
\begin{eqnarray}
&&\hspace{-0.3cm}\aleph(\gamma) = \bar{\alpha}_s\bigg(2\psi(1) - \psi(\gamma) \nonumber\\
&&\hspace{1cm} - \sum\limits_{n=1}^{N}{1\over n-\gamma}-\psi(N+1-\gamma)\bigg)\,,
\end{eqnarray} 
and considering that the BFKL limit in the Mellin space is ${\alpha_s\over \omega}\sim 1$,
the overlapping DGLAP-BFKL regime is achieved by further assuming $\alpha_s\ll\omega\ll1$. In the limit $\gamma\to 1$ we have 
$\aleph(\gamma) \to {\bar{\alpha}_s\over 1-\gamma}$ and 
${1\over \aleph(\gamma)-\omega}\to {1\over {\bar{\alpha}_s\over 1-\gamma}-\omega} = 
{1-\gamma\over \omega(\gamma -1 + {\bar{\alpha}_s\over \omega})}$. The first moving pole is obtained taking the residue at 
$\gamma=1-{\bar{\alpha}_s\over \omega}$ in eq. (\ref{quarkMellin1}).
The next-to-leading residue, \textit{i.e.} the next-to-leading moving pole of the expansion,
is at $\gamma=2-{\bar{\alpha}_s\over \omega}$. So, summing the first two residues we have
\begin{eqnarray}
&&\hspace{-0.2cm}
\int_{x^2_\perp M_N}^{+\infty}dL\, L^{-j}\, 
{1\over 2P^-}\langle P|\bar{\psi}(L,x_\perp)\gamma^-\nonumber\\
&&\hspace{-0.2cm}
\times[nL+x_\perp,0]\psi(0) |P\rangle
= {iN_cQ^2_s \sigma_0\over 24\pi^2\bar{\alpha}_s}
\,\left({Q^2_sx^2_\perp\over 4}\right)^{\!-{\bar{\alpha}_s\over \omega}}
\nonumber\\
&&\hspace{-0.2cm}
 \times\!\left(\!-{2\over x_\perp^2}{{P^-}^2\over M^2_N}+i\epsilon\right)^{\!{\omega\over 2}}
\left(1 + {Q^2_sx^2_\perp\over 5}\right)\,.
\label{q-2leadingres-sum}
\end{eqnarray}

The two leading residues we just calculated organize themselves as an
expansion in $x^2_\perp$: the sub-leading term is suppressed by an extra power of $x^2_\perp$ for $x^2_\perp\to 0$,
which is typical of the coordinate space twist expansion.

Let us observe that, in eq. (\ref{quarkMellin1}), besides the moving poles, we have also a non-moving pole
at $\gamma=1$.  Like we have seen in the gluon case~\cite{Chirilli:2021euj}, such pole is
canceled by diagrams that are not included in the high-energy OPE formalism.

Now that the first two leading twist contributions have been calculated, we can proceed with the 
calculation of the inverse Mellin transform. We will consider only the case $0<Q_s^2\Delta^2_\perp<1$ 
which is consistent with the twist expansion.
The inverse Mellin transform of (\ref{q-2leadingres-sum}) is
\begin{eqnarray}
	&&\hspace{-0.6cm}
	{1\over 2\pi i}\int_{1-i\infty}^{1+i\infty}\!\!\! d\omega\, L^{\omega}
	{iN_cQ^2_s \sigma_0\over 24\pi^2\bar{\alpha}_s}
	\,\left({Q^2_s|z|^2\over 4}\right)^{\!-{\bar{\alpha}_s\over \omega}}
	\nonumber\\
	&&\hspace{-0.6cm}
	\times\left({2\over z^2}{{P^-}^2\over M^2_N}+i\epsilon\right)^{\!{\omega\over 2}}
	\!\!\!\left(1 + {Q^2_s|z|^2\over 5}\right)
	\nonumber\\
	&&\hspace{-0.6cm}
	=  {iN_cQ^2_s \sigma_0\over 24\pi^2\bar{\alpha}_s}\!\!
	\left({4\bar{\alpha}_s\left|\ln{Q_s|z|\over 2}\right|\over \ln\left({2\nu^2\over z^2M^2_N}+i\epsilon\right)}\right)^\half 
	\!\!\!\left(\!1 + {Q^2_s|z|^2\over 5}\!\right)\! I_1(u) 
	\label{LT-NLTioffeamp}
\end{eqnarray}
where $I_1(u)$ is the modified Bessel function with
$u=\left[4\bar{\alpha}_s\left|\ln{Q_s|z|\over 2}\right|\ln\left({2\nu^2\over z^2M^2_N}+i\epsilon\right)\right]^\half$.

Result (\ref{LT-NLTioffeamp}) is the quark pseudo-ITD up to next-to-leading twist correction.
In Fig. \ref{Fig:IofAm-twistSadBFKLReIm}, we plot the real and imaginary parts of the 
leading (green dashed curve), and the next-to-leading (solid red curve), twist
corrections of eq. (\ref{LT-NLTioffeamp}), and compare them with the BFKL resummation result in the saddle point approximation,
eq. (\ref{q-saddpoint}), and the numerical evaluation of the integration over $\gamma$, eq.  (\ref{quark-bilocalresult-z}).
Note also that the curves for the LT and NLT corrections are almost one on top the other because the NLT is just a small shift with
respect to the LT.

{\it Quark pseudo-PDF at small-x$_B$}
---
In this section, we will perform the Fourier transform of the Ioffe-time distribution in the BFKL limit to get the quark pseudo PDF.
Then, we will do the same for the leading and next-to-leading twist corrections. Finally, we will compare
the pseudo-pdf in the BFKL approximation with the LT and NLT by plotting them in the range $x_B\in[0.01, 0.1]$.

Assuming $ 0\le x_B\le 1$, we have
\begin{eqnarray}
	&&\hspace{-1.3cm}\int_{-\infty}^{+\infty}\!\!{d\nu\over 2\pi}\,
	e^{-i\nu x_B}\calm(\nu,z^2)
	\nonumber\\
	&&\hspace{-1.3cm}
	= - {N_c Q_s\sigma_0 \over 8\pi^3|z|x_B}\int_{\half-i\infty}^{\half+i\infty} d\gamma
	\left({2\over z^2 M^2_Nx^2_B}+i\epsilon\right)^{{\aleph(\gamma)\over 2}}
	\nonumber\\
	&&\hspace{-1.3cm}
	~~~\times\!{\aleph(\gamma)}{\Gamma^2(1-\gamma)\Gamma^3(1+\gamma)\over \Gamma(2+2\gamma)}
	\left({Q^2_s|z|^2\over 4}\right)^{\gamma-\half}\,.
	\label{quark-pseudoPDF}
\end{eqnarray}
Result (\ref{quark-pseudoPDF}) gives the small-$x_B$ behavior of the quark pseudo-PDF with BFKL resummation. 
In saddle point approximation we have
\begin{eqnarray}
&&\hspace{-0.4cm}\int_{-\infty}^{+\infty}\!\!{d\nu\over 2\pi}\,
e^{-i\nu x_B}\calm(\nu,z^2)
= - {i\bar{\alpha}_sN_c Q_s\sigma_0 \over 32|z| |x_B|}{\ln 2 \over \sqrt{\chi}}
\nonumber\\
&&\hspace{-0.4cm}
~~\times\!\exp\left\{-{1\over \chi}\ln^2{Q_s |z|\over 2}\right\}
\left(\!{2\over x^2_Bz^2 M^2_N}+i\epsilon\!\right)^{\bar{\alpha}_s2\ln 2}\,,
\label{pseudoQpdfsaddle}
\end{eqnarray}
where we defined $\chi = 7\zeta(3)\bar{\alpha}_s\ln\left({2\over x^2_Bz^2 M^2_N}+i\epsilon\right)$.
The pseudo-PDF with BFKL resummation exhibits the usual resummation of $\ln 1/x_B$ with the exponentiation of the Pomeron intercept
$\bar{\alpha}_s2\ln 2$. 

Let us perform the Fourier transform of the pseudo-ITD with twist corrections, \textit{i.e.} we start from
eq. (\ref{q-2leadingres-sum}). The result is
\begin{figure}
	\begin{center}
		\includegraphics[width=2.6in]{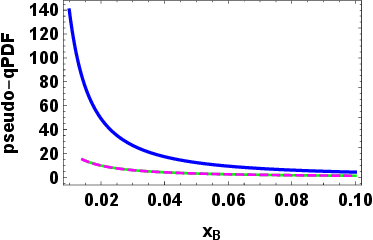}
		\caption{Plot of the quark pseudo PDF: we compare the saddle point approximation, eq. (\ref{pseudoQpdfsaddle}) (blue curve) and the LT (green dashed curve) and the NLT (magenta curve), eq. (\ref{qpseudoNLT}).}
		\label{Fig:Quarkpseudo-num-vs-sadd-vs-twist}
	\end{center}
\end{figure}
\begin{eqnarray}
&&\hspace{-0.6cm}
\int_0^{+\infty}\!\!{d\nu\over 2\pi}\,
e^{-i\nu x_B}\calm(\nu,z^2)
= {i\,N_cQs^2\sigma_0\over 48\pi^2 x_B}
\left(1 + {Q^2_s|z|^2\over 5}\right)
\nonumber\\
&&\hspace{-0.6cm}\times\!
{\ln\left(Q_s^2 |z|^2\over 4\right)\over \ln\left(-{2\over x_B^2 |z|^2M^2_N}+i\epsilon\right)}
\Big(J_0(m)-J_2(m)-{2\over m}J_1(m)\Big)
\label{qpseudoNLT}
\end{eqnarray}
with
$m\equiv \left[ 4\bar{\alpha}_s \left(\ln{Q_s|z|\over 2}\right)\ln\left(2\over |z|^2 x_B^2 M^2_N\right)\right]^\half$
and $J_0, J_1$, and $J_2$ the Bessel functions.
In Fig. \ref{Fig:Quarkpseudo-num-vs-sadd-vs-twist} we compared the pseudo-PDF with BFKL resummation,
eq. (\ref{pseudoQpdfsaddle}), and the LT and NLT, eq. (\ref{qpseudoNLT}). Note that, since the NLT correction is just a small
shift of the LT one, the respective curves are
one on top of the other. The pseudo-PDF (the following is valid also for the quasi-PDF)
have real and imaginary parts. The origin of the complex nature can be traced back to the phase in the evolution parameter, eq. (\ref{coordparam}).
Notice that, if we perform the Fourier transform integrating only over the positive
range, we get that the distributions are real, and their behaviors are the same.
In Fig. \ref{Fig:Quarkpseudo-num-vs-sadd-vs-twist},  we plotted only the real parts.

\begin{figure}
	\begin{center}
		\includegraphics[width=2.6in]{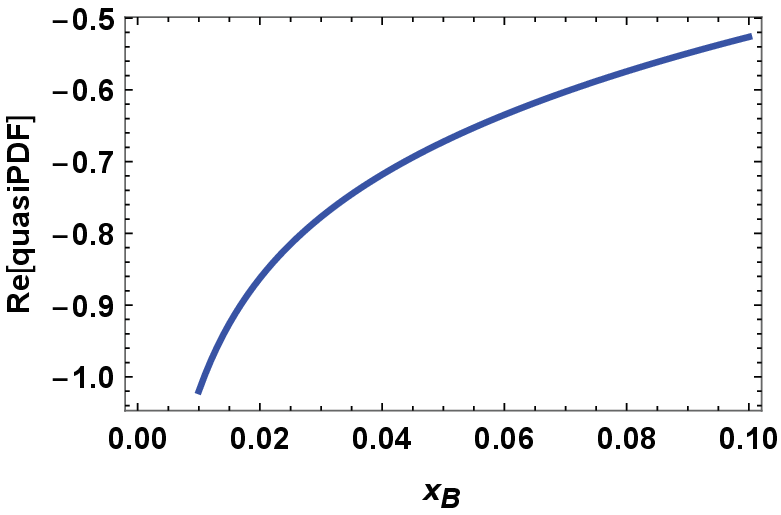}
		\caption{Plot of the quark quasi PDF, real part.}
		\label{Fig:quark-quasiPDFre}
	\end{center}
\end{figure}

{\it Quark quasi-PDF at small-$x_B$} ---
To obtain the quasi-PDF from the ITD we have to perform 
the Fourier transform with respect to $z^\mu$ keeping its orientation fixed. To this end,
we introduce a real parameter $\varsigma$ with $-z^2=\varsigma^2>0$, and define
the four-vector $\xi^\mu\equiv{z^\mu\over |z|} = {z^{\mu}\over |\varsigma|}$ with $|z|=\sqrt{-z^2}$.
Using $LP^- = |\varsigma|P_\xi$ with $P_\xi = \xi^\mu P_\mu$, the quasi-PDF Fourier transform is
\begin{eqnarray}
&&\hspace{-0.5cm}P_\xi\!\!\int_{-\infty}^{+\infty} \!\!{d\varsigma\over 2\pi} \, e^{-i\varsigma P_\xi x_B}
\calm(\nu,z^2) = {N_c P_\xi Q_s\sigma_0\over 4\pi}\int_{\half-i\infty}^{\half+i\infty} d\gamma
\nonumber\\
&&\hspace{-0.5cm} 
\times{\gamma^2\Gamma(\gamma)\over (4\gamma^2-1)\sin(\pi\gamma)}
\left(\!-{2P_\xi^2\over M^2_N}+i\epsilon\!\right)^{\!\!{N(\gamma)\over 2}}
\!\! \left({Q^2_s\over 4P^2_\xi x_B^2}\right)^{\!\! \gamma-\half}
\label{qfourier}
\end{eqnarray}
To finally obtain the quasi-PDF one should perform the last integration $\gamma$ which actually happens to be divergent at the point $\gamma=\half$. To overcome this issue we can first perform the integration over
$\gamma$ and leave the Fourier transform at the end. The numerical evaluation of the Fourier transform for the quark quasi-PDF is depicted in Fig. \ref{Fig:quark-quasiPDFre}. Upon examination, we observe that the distribution yields negative values, which corroborates the unsuitability of quasi-PDFs in the small-$x_B$ regime.

\begin{figure}
	\begin{center}
		\includegraphics[width=2.6in]{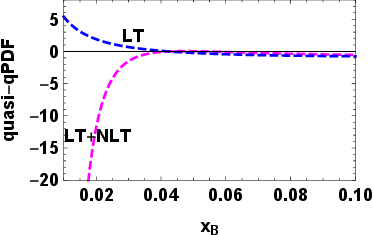}
		\caption{Plot of the leading and next-to-leading (blue and magenta dashed curve respectively) twist corrections of the
			quasi-pdf.}
		\label{Fig:Quarkquasi-LT-vs-NLT}
	\end{center}
\end{figure}

The quasi-PDF at leading and next-to-leading twist approximation is obtained 
using the Ioffe-time distribution at leading and next-to-leading approximation. So, using Fourier transform \`a la quasi-PDF
of the inverse Mellin transform of eq. (\ref{q-2leadingres-sum}), we have
\begin{eqnarray}
&&\hspace{-0.5cm}P_\xi\!\!\int_{-\infty}^{+\infty} \!\!{d\varsigma\over 2\pi} \, e^{-i\varsigma P_\xi x_B}
\calm(\nu,z^2) 
\nonumber\\
&&= {i\,N_cQ^2_s \sigma_0\over 24\pi^2 x_B}
I_0(t)\!\!\left(1 - {2Q^2_s\over 5P_\xi^2x_B^2}\right)
\label{NLtwistquasiPDF}
\end{eqnarray}
with $t \equiv \left[4\bar{\alpha}_s\ln\left({2P_\xi x_B\over Q_s}\right)\ln\left(-{2P_\xi^2\over M^2_N}+i\epsilon\right)\right]^\half$,
and $I_0$ the modified Bessel function.
We observe that, contrary to what was originally thought, 
for the quasi-PDF in eq. (\ref{NLtwistquasiPDF}) at small-$x_B$ the next-to-leading twist correction 
is enhanced rather than suppressed respect to the leading one. 
In Fig. \ref{Fig:Quarkquasi-LT-vs-NLT} we plotted only the real part of the quasi-PDF. 

{\it Conclusions} ---
We showed that
the high-energy behavior of the pseudo Ioffe-time distribution is governed by the resummation of 
the large logarithms of the Ioffe-time parameter $\nu=z\!\cdot\!P$ which acts like the evolution parameter of the BFKL equation
(see Fig. \ref{Fig:IofAm-twistSadBFKLReIm}).
We calculated the pseudo-PDF with BFKL resummation in the
saddle point approximation, eq.  (\ref{q-saddpoint}) 
(see Fig. \ref{Fig:Quarkpseudo-num-vs-sadd-vs-twist}) and showed that it has the expected small-$x_B$ behavior 
characterized by the famous Pomeron intercept at leading-log approximation.
The quasi-PDF, instead, turned out to be negative at this regime (see Fig. \ref{Fig:quark-quasiPDFre}).
Performing an analytic continuation from integer to non-integer spin, we obtained the leading and next-to-leading 
twist corrections to the structure functions defined through the pseudo-ITD, eq. (\ref{defoperator}): we calculated the 
leading and next-to-leading residues of eq. (\ref{quarkMellin1}), which are the first two terms of an infinite series whose entire resummation coincides with the BFKL result.  
Similarly to what was observed
in the gluon case~\cite{Chirilli:2021euj}, in the pseudo-PDF case, the next-to-leading twist correction does not seem to
be approaching the full BFKL behavior much faster than the leading twist one, as it is commonly believed. Indeed, as can be observed from Fig. 
\ref{Fig:IofAm-twistSadBFKLReIm}, the next-to-leading twist correction appears to be just a small shift. This means that
the true small-$x_B$ behavior can only be captured from the full BFKL result and that the twist corrections at small-$x_B$ are
irrelevant unless the full series of twist corrections is taken into account. We can conclude that the analytic structure of 
each term of the series of twists is different from that of the sum of the series \textit{i.e.} the BFKL resummation.
For the quasi-PDF case, instead, the first two twist corrections exhibits
rather unusual behavior at small-$x_B$ (see Fig. \ref{Fig:Quarkquasi-LT-vs-NLT}).

{\it Acknowledgments} ---
The author is grateful to Ian Balitsky and Vladimir Braun for valuable discussions.

\bibliographystyle{JHEP}
\bibliography{/Users/chirilli/Documents/mm/m/MyReferences}

\end{document}